# Ancient and modern rangefinders

**Amelia Carolina Sparavigna**
Department of Applied Science and Technology
Politecnico di Torino, C.so Duca degli Abruzzi 24, Torino, Italy

*Rangefinders are instruments used for ballistics and for surveying in general. Here a short discussion of some of them, ranging from the ancient Rome to the modern methods.*

OAKOC is an Army acronym used for planning operations: it stands for Observation and fields of fire, Avenues of approach, Key and decisive terrain, Obstacles, and Cover and concealment. The first point is therefore the Observation and fields of fire. It means that it is necessary to answer the following questions [1]: What can you see throughout your area? What fields of fire will you have in terms of maximum effective range? How will the weather affect your and enemy's visibility?

First of all, it is required a surveying to measure the ranges of location and fire. Here the noun "range" has the two following senses. It means the extent or limits between which an operation or action is possible. It has also the sense, pertaining to ballistics, which means the distance to which a projectile is or may be sent by a weapon or the distance of the target from the weapon [2].

For what concerns the etymology, "range" as "row or line of persons," hunters or soldiers, is coming from the Old French "rangier, to place in a row, arrange." Meaning also "scope, extent" was first recorded 1660s. The sense of "distance a gun can send a bullet" is recorded from 1590s. Therefore, the "ranger" is the agent noun from range (verb), in the modern military sense of "member of an elite combat unit" [3].

An instrument is a "rangefinder," if it is suitable for determining the distance from the observer. It is therefore necessary for OAKOC. In fact, the noun rangefinder, is used for sighting a gun and also for adjusting the focus of a camera. According to the Websters dictionary [4], it is a measuring instrument (acoustic or optical or electronic) for finding the distance of an object. In gunnery, an instrument, or apparatus, variously constructed, for ascertaining the distance of an inaccessible object, used to determine what elevation must be given to a gun in order to hit the object.

Here we will discuss rangefinders, choosing some devices, ranging from the ancient Rome to the modern instrumentation.

**1. An ancient rangefinder (the Roman Dodecahedron)**
It is easy to guess that OAKOC is a procedure as old as war. To see how Julius Caesar managed OAKOCs we can follow some precise descriptions that he wrote in his De Bello Gallico. In the English translation [5] of this book, we find the noun "range" used in the sense of prepare a "row or line of soldier." The term "distance" is used in the English translation for an estimate of the physical separation of Roman soldiers from enemies. Caesar gives known distances in miles, but when he reports the fights he describes the relative ranges telling for instance, "a short distance from the legions," "small distance," "great distance" and "very great distance." We could ask ourselves, are these terms simply proposing an estimate or are coming from some measurements?

After studying some Roman artifacts and the Latin texts by Caesar and other writers, scholars have provided some discussions of ancient surveying methods. Romans used "groma" and the "cross" as surveying instruments [6]. As told in Ref.7, the surveyor's cross is a simple instrument for making alignments. Its primary feature is the use of some vertical slits, positioned opposite and at right angles to each other, which are inches or feet apart, depending on the form of the instrument. "By

lining them up, lines can be projected on the ground for a considerable distance." [7]

For fast planning operations of an army or for launches of ballistic machines, had the Roman any different suitable instrument? A possibility is the use of a Roman Dodecahedron [8,9]. Fig.1 on the left shows one of them. These dodecahedra are bronze artifacts dating from the 2nd or 3rd century AD. These objects always consisting of 12 regular pentagons and have a diameter ranging from 4 to 11 centimetres. Several of them have at the centre of the 12 faces, holes of different sizes.

It is usually told that the Roman dodecahedra, about one hundred collected in several European museums, came from Gaul and the lands of the Celts. In fact, the lands where these artifacts had been found were at the border of the Empire. It means that the dodecahedra were probably connected with military operations. According to [9], their function or use is considered as a mystery. Moreover there are so many theories reported by the World Wide Web (candlestick holders; dice; survey instruments), often without a proper reference, that it is impossible to verify them. The hypothesis of using of them as dice is weak, because a roman dodecahedron such as that in Fig.1 is biased, and therefore not suitable for sorting or gambling (dodecahedral dice were quite different [10]).

I have prepared a dodecahedron, made of paper, according to the data of one of them given in Ref.11. In this reference is reporting the sizes of the holes (see Fig.1 on the right). The dodecahedron, found at Jublains, the ancient Nouiodunum, is dating from the $2^{nd}$ or 3rd century AD. In [8], I proposed this Roman object for measuring distance, using a method based on similar triangles. The dodecahedron possesses five angles of view, and, knowing the size of an object allows to determine its distance. The angle of view is the cone which describes the angular extent of a given scene.

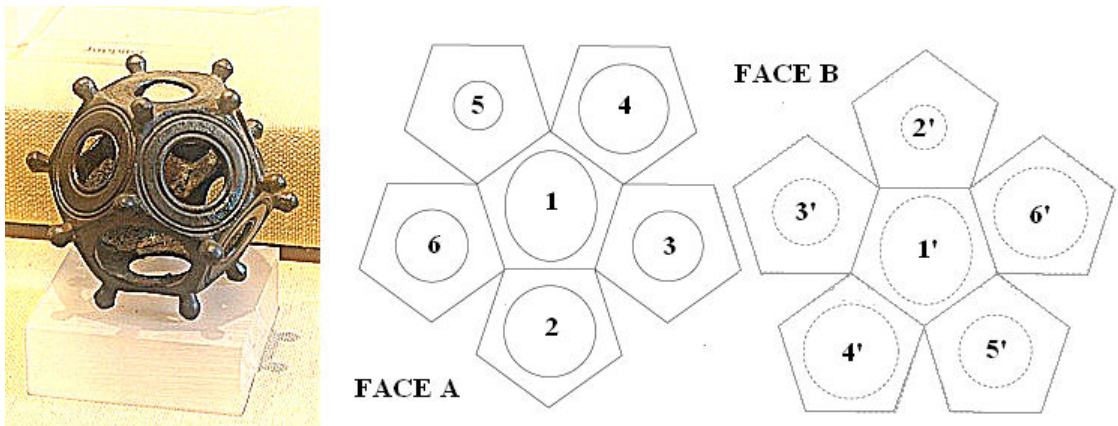

**Fig.1 On the left a dodecahedron (source: Wikipedia). On the right the faces of the roman dodecahedron of Ref.11.**

Naming 1,2,3,4,5, and 6, the holes of face A, and 1',2',3',4',5' and 6' the holes, of face B (see Fig.1, we have the following opposite pairs: (1',1),(2',6), (3',5), (5',3), (4',4) and (6',2) (for they size see Ref.8). Let us consider for instance, the pair (2',6) and look through the dodecahedron, holding it with 2' and 6 parallel, with 2' near an eye and 6 as opposite. If the dodecahedron is close the eye, we see the two holes (see Fig.2); if it is too far, we see just the nearest hole 2'. There is a distance where we can see the circumferences of the two holes as perfectly superimposed. This gives the precise angle of view, that we can use for measurements.

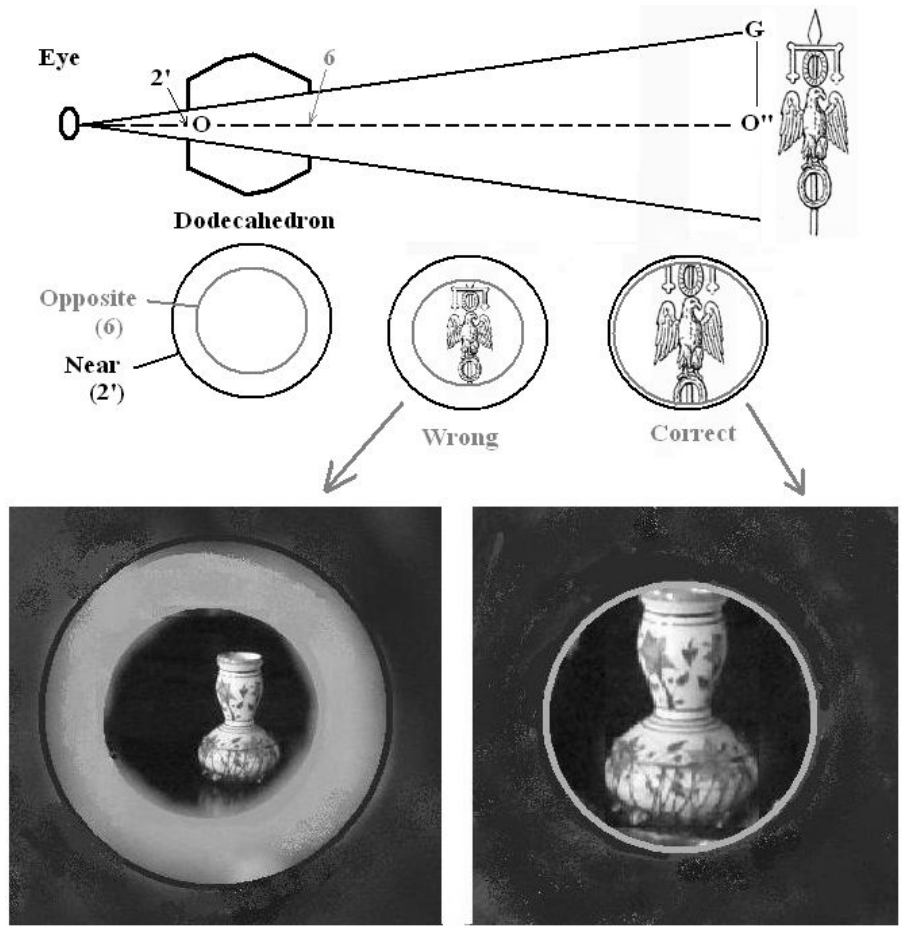

**Fig.2** Let us consider the pair (2', 6) and look at a Roman vexillum through the dodecahedron, holding it with 2' and 6 parallel, with 2' near an eye and 6 opposite. If the dodecahedron is close the eye, we see the two holes. If it is too far, we can see just hole 2'. There is a distance where we can see the circumferences of the two holes (black and grey in the image) as perfectly superimposed. This is a specific cone of view that we can use for measurements. In the lower part of the figure, a "home version" of the sketch, obtained with the pair (3',5) of a dodecahedron made of paper. The vase is 0.2 m high and the dodecahedron at a distance of 2.5 m from it.

Points O,O'' and G are given in Fig.2 (for more details, please see Ref.8). OO'' is the eye-line. After the sizes of holes, we have for the pairs of the dodecahedron the following ratios:

| Pair | GO''/OO'' |
| --- | --- |
| (2',6) | 0.065 |
| (3',5) | 0.04 |
| (4',4) | 0.01 |
| (5',3) or (6',2) | 0.005 |

let us use the following example to show how to measure distance. A Roman soldier is looking through the dodecahedron at a target, for instance a vexillum, that he knows it has a height of two meters. If the soldier sees the vexillum fitted in the field of view of the pair (2',6), that is, with its height coincident with the diameter of the superimposed circumferences 2' and 6, what is its distance? Looking at the table, for the vexillum 2 meters long, GO''=1 m, then OO'' ≈15 m. This is the result for pair (2',6). Using the other pairs, we have distances of 25 m, 100 m and 200 m, respectively. Then the soldier can use the device for four different ranges.

In the case of the small vase shown in the lower part of Fig.2, which is 0.2 m high, I used the pair (3',5). The dodecahedron was at 2.5 meters. I observed the two circumferences as superimposed and the object perfectly fitted in the hole.

Of course it is not necessary that the target is perfectly fitting the hole. It is necessary that the soldier, looking through the dodecahedron, is able to estimate the size of the scene containing the target. In this way, he will be able to evaluate an approximate distance of it. For me, those dodecahedrons having a structure with holes of different sizes, are military instruments to evaluate distances for ballistics. Roman dodecahedron is a rangefinder simple to use.

## 2. The fore-staff

The previously discussed dodecahedron, being of bronze was quite strong and portable. Considering it as a rangefinder of the Roman Army, its use was lost after the collapse of the Empire. It seems that the Middle Age developed a different rangefinder, or used an already known device. It was the fore-staff. The noun "fore-staff" is referring to several things and therefore it is a little bit confusing. The best is to describe it according to its use. It is also known as Jacob's staff or cross-staff when it is used for astronomy and navigation, as a device to measure angles, later replaced by sextants. In surveying, it is a vertical rod that penetrates the ground, supporting a compass for measuring angles. I am using "fore-staff" because it is the best for describing the device used in ballistics. As told in Ref.12, the fore-staff is also known as "ballista." In fact, the ballista is an ancient military engine, which was used for throwing stones, darts and javelins. Therefore the rangefinder used to determine the distance of the target was named after its ballistic machine.

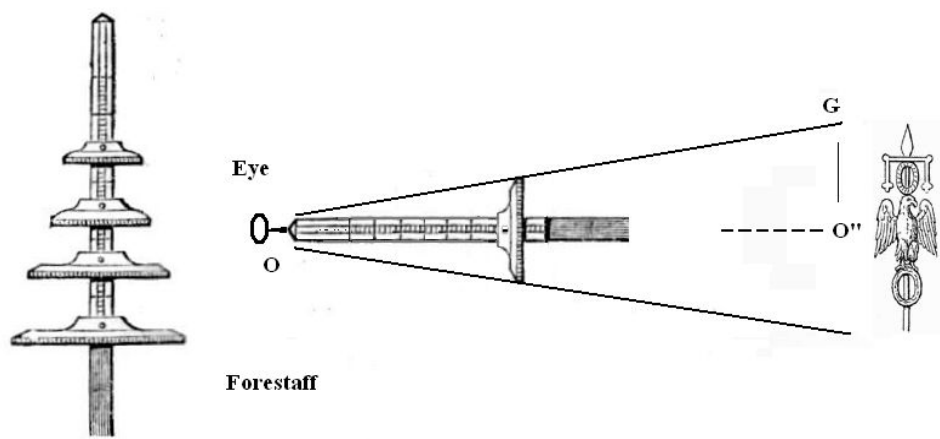

**Fig.3 Fore-staff and how to use it on a eye-line.**

As we can see the Figure 3, it is quite simple and its use based on similar triangles again. In the original form of the fore-staff, it is a pole (the main staff) marked with graduations for length. A cross-piece, called transom or transversal, slides on the main staff. It was common to provide several transoms, each with a different range of angles it would measure.

This rangefinder is quite large, moreover it has the drawback that one needs to change the transom for targets at different distances. The dodecahedron is small, and to change the range, it is enough to rotate it. It has the same features as the fore staff, but is more convenient for military actions.

### 3. The coincidence rangefinder

The modern rangefinders are used for determining the focus in photography, or again, as we have discussed before, for accurately aiming a weapon. Some instruments are based on active methods, using emission of energy by means of sonar, laser, or radar. The laser rangefinder can operate on the time of flight principle. They launch a laser pulse towards the object and measure the time this pulse needs to come back to the instrument. Moreover, using the Doppler effect it is possible to measure how fast is the target moving (these devices in Italy are known as Autovelox).

However, the use of optical instruments based on triangulation methods persists (stadiametric rangefinders, parallax, coincidence rangefinders) [13]. These optical methodologies have been in regular use since the 18th century. Ref.14 reports that the coincidence range finder uses a single eyepiece. It is more or less a cylinder. Light from the target enters through two windows spaced apart on the cylinder. The distance between these windows is the base length B of the rangefinder. The incident beams are reflected to the center of the optical cylinder by prisms. to form two images of the target which are viewed by the observer through the eyepiece. Since the beams enter the instrument at slightly different angles, an observer sees a blurry image. Adjusting a compensator, the observer can tilt one of the beams to match the two images. In such a manner, the images are "in coincidence." The rotation of the compensator gives the distance of the target by triangulation.

Ref.15 is proposing a simple coincidence instrument (see Fig.4).

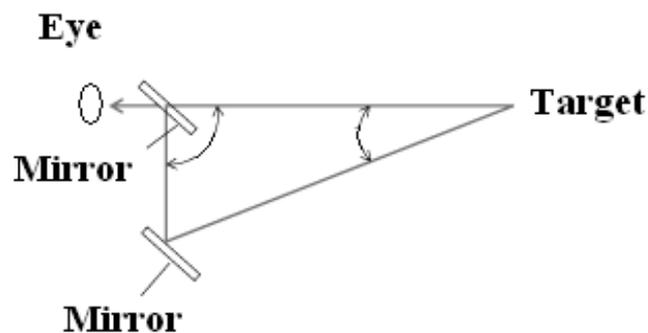

**Fig.4 Simple coincidence rangefinder. The two angles are a right angle and the convergence angle A at target.**

It is told in [15] the rangefinder operates as an angle-measuring device, using the triangle comprising the rangefinder base length B and the line from each window to the target point. The basic optical arrangement is shown in Fig.4, where the mirror in front of the eyepiece is semitransparent. When we have the coincidence, the target T is seen in the same apparent position. We have that the rangefinder equation is satisfied. The distance D from the target is: $D = B \cot A$.

The fact that we need to have a coincidence of images, is, more or less the same condition we need for the use of a Roman Dodecahedron, where we have to find the coincidence of the opposite hole circumferences. Therefore, we could conclude telling that it was the Roman coincidence rangefinder.